\documentclass[showpacs,preprintnumbers,amsmath,amssymb]{revtex4}
\usepackage{graphicx}
\usepackage{dcolumn}
\usepackage{bm}
\usepackage{tensor}

\begin{document} 
\title{Two-dimensional periodic and quasiperiodic spatial structures \\
in microchip laser resonator.}
\author{A.Yu.Okulov}
\email{alexey.okulov@gmail.com}
\homepage{https://sites.google.com/site/okulovalexey}
\affiliation{Russian Academy of Sciences, 119991, Moscow, Russia}

\date{\ September 22, 2002}
  
\begin{abstract}
{ The spatially periodic 2D patterns at output mirror 
of solid state microchip laser with high Fresnel number (100-1000)
are discussed in view of numerical modeling 
with split-step FFT code comprising nonlinear gain, 
relaxation of inversion and paraxial diffraction. }
\end{abstract}

\pacs{42.65.Hw,42.65.Jx,42.65.Re,42.55.Wd,42.60.Jf}

\maketitle
 
\section{Introduction}
 
 Spatially periodic structures of electromagnetic field in optical 
cavities could arise not only due to boundary conditions, 
as for example in the case of rectangular 
waveguide \cite {Weinstein:1969}. 
In Talbot cavity \cite {Napartovich:1987}  the  spatially periodic layout of 
cavity parameters forces the lightwaves to follow the profile 
of index and gain. The more interesting situation occurs when 
nonlinear wave interaction itself arranges the 
sophisticated electromagnetic structures  
\cite {Napartovich:1987_jtpl,Okulov:1990,Firth:1992}. 
Regardless to the physical 
nature of  nonlinearity the common feature of these structures 
is their translational symmetry: in passive systems the hexagonal 
spatial structures are dominant \cite {Napartovich:1987_jtpl,Firth:1992},
in active systems having 
optical gain the rectangular structures are more likely to 
survive 
\cite {Napartovich:1987_jtpl,Okulov:1990,Staliunas:1995}. 
The map of parameters space for each given 
system contains regions with hexagonal, rectangular arrays, 
spatial localized structures (spatial solitons or diffractive 
autosolitons \cite{Okulov:1988,Rozanov:1997,Likhansky:1993,Lugiato:1998}) 
and spatial chaos \cite {Okulov:1988,Hollinger:1985}. 
The location of these regions and their boundaries 
sensitively depends upon geometry of optical cavity, i.e. on positions 
and curvicity of mirrors, lenses, nonlinear elements and apertures.
Of course, 
in real experimental practice \cite {Chen:2001,Kubota:1992} 
such separation is often 
ambiguous, because almost any optical element, for example gain 
element, could  have properties of lens (linear or nonlinear), 
partially reflecting mirror, aperture,  birefringency et al. 
Nevertheless, recent experimental results show the stable 
electromagnetic field patterns \cite {Chen:2001,Kubota:1992} described by 
relatively simple and robust theoretical models.

This models are reduced from conventional  Maxwell-Bloch 
equations for two-level gain medium \cite{Eberly:1975}, 
nonlinear wave equation for $\chi^2$ (parametric ) \cite {Sukhorukov:1976}, 
$\chi^3$ ( Kerr ) \cite {Newell:1990},
 or photorefractive media \cite{Soskin:1991}. 
We will restrict here ourselves 
by two-level resonant nonlinearity which is conventional basis 
for description of the laser dynamics \cite{Oraevsky:1964}. 
The most interesting feature of dynamics observed in  both  
numerical  modeling  and experiments is the possibility of reducing the 
entire set of Maxwell - Bloch  equations \cite{Eberly:1975,Oraevsky:1964} 
to equivalent 
single evolution equation:  Ginzburg - Landau  
equation \cite{Okulov:1988,Collet:1989}
or  to  the  more complicated  Swift-Hohenberg  equation 
\cite{Hohenberg:1993,Staliunas:1998}. 
In the Swift-Hohenberg approximation the  additional 
diffusion - like terms arise in the equation for optical 
envelope, because finite bandwidth of atomic gain line is 
taken into account and consequently, the dissipative 
filtering of  higher harmonics.  Such  reducing  works especially well  
in class - A laser,  when  relaxation $T_1$ ,  $T_2$   of atomic variables $N$ , 
$P$ (atomic inversion and polarization)  is significantly  faster  
than  relaxation $\tau$ of electric field 
envelope  $E$ :  $\tau << T_1  ,T_2$ . 
In class - B laser \cite {Staliunas:1995,Chen:2001,Kubota:1992}
the approximation of single 
evolution equation for electromagnetic field is insufficient,
 because the  fast damping of  optical envelope $T_1 <<\tau<<T_2$, 
leads to relaxation oscillations caused by slow recreation of 
population inversion \cite {Staliunas:1995,Hollinger:1985,Chen:2001}.  
In  this case  the additional equation for 
population inversion added, and the most elaborated approach 
is in  inclusion of the Swift-Hohenberg's diffusive 
terms \cite {Staliunas:1995}. 
In general these nonlinear systems 
could have about  
$\Delta N = V \nu^3 \Delta \nu /c^3$ modes comprised within 
 cavity volume $V$ and frequency interval, 
but nonlinear dynamics  led to spontaneous transverse 
mode-locking described firstly in \cite {Auston:1968}. 
such a way  that as it had been restricted in choice of possible modes.

Our goal here is in developing the alternative approach which 
uses integral 
equations rather than differential ones. In fact it is the 
experimental feature of 
solid state laser microcavities that radiation flight - 
time between mirrors is faster 
than all typical times in Maxwell-Bloch 
equations   $2L / c  <<  T_1, T_2,\tau$. Radiation 
lifetime in cavity is defined usually as $\tau = 2L/c(1-R)$. 
Thus it is possible to consider 
atomic variables $N$, $P$ as "frozen", 
while radiation $ E$ passes through  gain medium. Then 
this variables are forced by constant  external field $E$ (optical field) , 
while the 
latter bounces between mirrors \cite {Hollinger:1985}. 
Such separation leads to discrete mapping of 
intracavity field from one bounce to another, 
the number of bounce  $n$ serves here as 
discrete time $t$ . This approach proved  
to be fruitful for both the class - B laser 
(ring Nd-YAG long cavity) with low Fresnel  number, 
when only several transversal modes 
excited \cite {Hollinger:1985} and high Fresnel-number microchip 
laser with short cavity \cite {Okulov:1990}. In this case 
the spatiotemporal evolution of optical 
envelope is governed by iterative mapping of 
convolution type \cite {Okulov:1990,Okulov:1988,Hollinger:1985}: 

\begin{eqnarray}
\label{conv_map}
E_{n+1}(\vec{r_2} )= \int  K(\vec{r_1}-\vec{r_2}) f(E_{n}(\vec{r_1})) 
d \vec{r_1},
&& \nonumber \\
N_{n+1} =-{\frac{N_{n}- N_{0}}{T_1}}-{\sigma}N_{n}{\:}|{E_{n}}|^2,
\end{eqnarray}
where $\sigma$ is stimulated emission 
cross-section \cite {Hollinger:1985}.

The map (\ref {conv_map}) is $nonlocal$ , i.e . it introduces 
 the severe spatial as well as 
temporal  dispersion at each iterate (radiation bounce). 
In the above mentioned mapping 
 set  the nonlinearity action $f$, which usually acts as 
sharpen the field distribution, 
is described by local nonlinear mapping  
$f$, while convolution with kernel $K$ introduces 
the spatial dispersion. The most interesting feature of 
such approach proved to be even 
more  general in some sense than starting 
 point: Maxwell - Bloch equation for class B 
- laser, received by naive adiabatic elimination of 
polarisation field  $P$ \cite {Likhansky:1993,Lugiato:1998}. 
The case  is  that  boundary conditions are included directly 
into the convolution integral 
and, consequently, spatial  filtering  provides  
the diffusive terms  \cite {Okulov:1988} which in 
general are  not restricted to second order. 
This fact leads to the following results: 

1. Nonlocal iterative map (\ref {conv_map}) contains both 
Ginzburg - Landau and  Swift - 
Hohenberg equations. The latter arise even in the case of 
adiabatic elimination of 
polarisation field $P$.  

2.	The existence of spatially periodic structures which are 
"fixed points" of the 
map (\ref {conv_map}) or stable solutions of the 
Ginzburg - Landau or  Swift - Hohenberg equations  
is  manifestation of Talbot phenomenon, when spatial period of translationally 
symmetric structures is self-chosen in such a way to fulfill the condition of 
self-imaging . 

3.	Experimentally obtained  spontaneously arising  vortex  array structures  in 
microchip laser resonator have period such that Talbot condition 
on cavity length $L = 2P^2/\lambda $
fulfilled  and Fresnel number $Fr$  of cavity  proved  to be  in the 
vicinium of the value $Fr = N_v^2$   , predicted in \cite {Okulov:1990}. 

\section {Different  forms of the evolution equations}  
\subsection 	{ Partial  differential  equation  form}
 
The starting point of current analysis are 
Maxwell - Bloch equations \cite {Weinstein:1969,Eberly:1975,Oraevsky:1964},
written for slowly varying amplitudes and in paraxial approximation: 

\begin{eqnarray}
\label{Maxw_Bloch}
\frac {\partial E}{\partial z} + \frac {1}{c} \frac {\partial E}{\partial t} 
+ \frac {i}{2k} \Delta_{\bot} E = -\frac { i 2 \pi \omega}{c} P - \gamma E
&& \nonumber \\
\frac {\partial N}{\partial t}= \frac { N_0 - N}{T_1} - \frac {2i}{\hbar} 
(E P^{*} - P^{*}E),
&& \nonumber \\
\frac {\partial P}{\partial t} = -{\frac{P}{T_2}}- i (\omega -\omega_a)P +
\frac {i \mu^2}{\hbar} {E N},
\end{eqnarray}
where $E$ , $N$ , $P$ are optical field , 
inversion and polarization of the medium (density 
of the dipole moment of the resonance impurity 
in solid-state dielectric) respectively,
$\mu$  is electric dipole moment, $k = 2\pi /\lambda$ , $c$ 
- speed of light, $\gamma$  are  nonresonant losses in medium, 
$\omega_a$ is atomic resonance frequency, $\omega$ is the carrier 
frequency of the 
optical field $E$.  

The spontaneous emission sources are neglected 
and polarization of electromagnetic 
field (the mutual orientation of electric 
 and magnetic fields with respect to 
wavevector $\bf k$) is assumed here to be homogeneous 
over the entire aperture.  
	The further simplification of MBE's (\ref {Maxw_Bloch}) could be 
 fulfilled traditionally 
\cite {Newell:1990,Oraevsky:1964}  by virtue of longitudinal averaging of  
$E$ along  optical  axis  $z$ over cavity length $L_r$ :
 
\begin{eqnarray}
\label{Maxw_Bloch_averaged}
\frac {\partial E}{\partial t} 
+ \frac {i}{2k} \Delta_{\bot} E = - i 2 \pi \omega  P - \frac {E}{\tau}
-i (\omega_c -\omega ) E 
&& \nonumber \\
\frac {\partial N}{\partial t}= \frac { N_0 - N}{T_1} - \frac {i}{2\hbar} 
(E P^{*} - P^{*}E);
&& \nonumber \\
\frac {\partial P}{\partial t} = -{\frac{P}{T_2}}- i (\omega -\omega_a)P +
\frac {i \mu^2}{\hbar} {E N},
\end{eqnarray} 
This is  so-called  C - class  laser  MB   equations.  
The next  reduction  could be 
fulfilled by two ways . The first  one  is  in  trivial  elimination  of  the  
polarisation  variable $P$  under "obvious " condition : 
\begin{equation}
\label{atomic incoherency}
\frac {\partial P}{\partial t} << \frac {P}{T_2} 
\end{equation} 
which  leads  to  the  quasistatic elimination of P in  the form : 
\begin{equation}
\label{qausistatic polarization}
P=\frac {i \mu^2 E N}{\hbar (i(\omega -\omega_a )T_2+1) }. 
\end{equation} 

After substitution of $P$ into (\ref {Maxw_Bloch_averaged}) 
the MBE system becomes : 
\begin{eqnarray}
\label{Maxw_Bloch_incoherent}
\frac {\partial E}{\partial t} 
+ \frac {ic}{2k} \Delta_{\bot} E = \sigma E N - \frac {E}{\tau_c}
-i (\omega_c -\omega ) E,  
&& \nonumber \\
\frac {\partial N}{\partial t}= \frac { N_0 - N}{T_1} - \sigma |E|^2 N; 
\sigma = \frac {2 \pi \mu^2 \omega T_2}{\hbar c (1+i(\omega -\omega_a))}
\end{eqnarray} 
the so-called  B - class  laser  MB   equations. 
The  system  ( \ref {Maxw_Bloch_incoherent} ) has  
intermediate 
time scale , the relaxation  oscillations  time :  
\begin{equation}
\label{relaxation period}
\tau_{rel}= \sqrt{T_1 \tau_c} / \sqrt {\sigma N_0 c \tau_c-1}, 
\end{equation} 				
whose  value  depends   upon  relative  values   of  times $T_1, \tau_c$ ,
 determined by the 
physical  nature  of  gain medium . 
In solid - state  medium, such as rare - earth 
doped dielectrics, the  $\tau_rel$  lies  
in  between $T_1<<\tau_rel<<T_2$.  In  the 
 limit $\tau_rel>> T_1,\tau_c$,    i.e.  
just  near  lasing threshold, 
 when  radical  in  (\ref {relaxation period}) is  very small,   
the  system (\ref {relaxation period})
is  turned  into system for class - A  laser : 
\begin{equation}
\label{class_A_laser}
\frac {\partial E}{\partial t} 
+ \frac {ic}{2k} \Delta_{\bot} E = 
\frac {\sigma E N}{2 (1+\sigma T_1 |E|^2)} - \frac {E}{\tau_c}
-i (\omega_c -\omega ) E,
\end{equation}  
which  has the  form  of  the Ginzburg - Landau equation 
\cite {Okulov:1988, Rozanov:1997, Newell:1990,Collet:1989,
Staliunas:1998} . 

The  more  rigorous approach  had  been developed
in \cite {Staliunas:1993,Newell:1994}, 
where it was found that polarisation  of  medium  provides  
natural  spatial filtering , owing  to  finite 
linewidth  : 
\begin{eqnarray}
\label{Swift_Hohenberg_filtering}
\frac {\partial E}{\partial t} 
+ \frac {ic}{2k} \Delta_{\bot} E = \sigma E N - \frac {E}{\tau_C} 
-i (\omega_c -\omega ) E + 
&& \nonumber \\
T_2^2 [(\omega_a-\omega)+\frac{ic}{2k} \Delta_{\bot}]^2,  
&& \nonumber \\
\frac {\partial N}{\partial t}= \frac { N_0 - N}{T_1} - \sigma |E|^2 N; 
\end{eqnarray} 

This complex  Swift - Hohenberg 
equation \cite {Staliunas:1995,Staliunas:1993,Newell:1994}, 
where dissipative terms, 
containing  second  and  fourth  order  spatial  derivatives  provide effective  
smoothing of abrupt changes of  spatial  structure  of  electromagnetic field and  
suppression of collapsing instabilities, 
typical  to  both  Ginzburg - Landau equation  
and  nonlinear Shrodinger equation  \cite {NN:1986}.  
These  terms  are  disappearing  in  the 
limit  $T_2 \rightarrow 0$  ,  i.e. when  atomic  linewidth  
tends  to  infinity. 

\subsection 	{ Integral  equation  form}

	The main result of the current paper is that such 
spatial filtering had been 
already included in  familiar model  of laser  
dynamics, elaborated in \cite {Okulov:1988}, which had 
been applied  afterwards to numerical simulations of  optical 
field structure of the 
class - A laser with spatially periodic gain 
distribution \cite {Okulov:1993,Okulov:1991,Okulov:1991spie}. 
The idea of inclusion 
the spatial filtering directly into the evolution 
equation of electromagnetic field is 
not new. This idea belongs to Fox and Li \cite {Fox_Lee:1966} 
whose  model  consists of the infinite  
sequence of  periodically located nonlinear amplifying screens , 
which imitates the 
radiation  bounces  from one mirror of the laser resonator to another. 
Boundary conditions are taken into account 
 by multiplying the kernel $K$ by both amplitude-phase 
masks  which imitate reflections from  mirrors,
 limitations by edges of cavity elements 
etc. In fact, the Fox - Lee  method  is no more than discrete mapping  
the two-dimensional  complex field  from  one mirror  
to another  by  means of a product of 
two consecutive mappings : 

one is local  and  nonlinear : it  acts  on each point of spatial structure 
independently of it's neibourghs, 
the other is  nonlocal and  linear  :  it mixes the mutual  actions of  
adjacent points with each other 
This procedure had been used each time, 
when robust and efficient computational  scheme 
had been required \cite {Sziklas:1975}. T
he spatial filtering had been introduced in split - step FFT 
method as "windowing" and "sampling" and  the  
authors had  not loose time in vain 
attempts to mask  the analogies with Fox-Lee 
method \cite {Hollinger:1985, Fox_Lee:1966,Sziklas:1975}. 

	It was shown in \cite {Okulov:1988} that infinite sequence 
of alternating nonlinear amplifying  slices and  spatial  filters 
 could  be modelled  by  nonlocal map with $real$ $kernel$  
when  confocal cavity of the so-called $8F$ - type considered:
\begin{equation}
\label{conv_map2}
\ E_{n+1}(\vec{r_2} )= \int  K(\vec{r_1}-\vec{r_2}) f(E_{n}(\vec{r_1})) 
d \vec{r_1}.
\end{equation}
	In  same way the evolution  map for radiation in optical 
cavity with arbitrary 
curvature of mirrors (fig.1) obtained  : 
\begin{eqnarray}
\label{complex_8F}
 E_{n+1}= \hat {Fr}  {\:}  f [E_n] {\:} {\:} {\:} 
, \hat {Fr}  {\:}  f [E_n(\vec r_{\bot})]= 
&& \nonumber \\
 \frac{i k}{2 \pi L}{\:}
{\int_{-\infty}^\infty}{E_n(\vec r_{\bot}^{'})} 
\exp[\frac{ik(\vec r_{\bot}-\vec r_{\bot}^{'})^2}{2L}]
{D(\vec r_{\bot}^{'})} 
d (\vec r_{\bot}^{'}),
\end{eqnarray} 
where  $\hat {Fr}$ is the solution of parabolic 
wave equation ( linear Shrodinger equation ) over  
cavity length L  in standard convolution form
 ( the kernel is  Green function ), $ D$ 
is complex aperture function \cite {Okulov:1991} whose  
modulus  corresponds to finite width of  gain 
medium, mirrors, lenses  etc. , while argument  imitates  
the phase modulation  
produced  by  curved  surfaces with effective focal  
length  $F$  and  random phase  
modulation  produced by  random  field  $\psi$  of  the  roughness  :
\begin{equation}
\label{complex_aperture2}
D(\vec{r_{\bot}} )= D_o(\vec{r_{\bot}} )
\exp (i \frac{k \vec r_{\bot}^2}{2F} + i \psi (\vec r_{\bot}))
\end{equation}
 
The equations ( \ref {complex_8F} - \ref{complex_aperture2} )  
are  the rigorous  \cite {Okulov:1990,Okulov:1988,Hollinger:1985}
 for  intracavity  field 
dynamics  during  the single radiation round - trip for   
class - A  laser , provided  
the gain medium and phase 
inhomogeneities are concentrated nearby the given plane of the 
cavity (fig.1). Otherwise the more complicated version 
of the equations ( \ref {complex_8F},\ref{complex_aperture} ) is 
used \cite {Sziklas:1975} , when gain medium 
is represented as a sequence of thin slices, each with its 
own gain and phase  ripple. The iterates  of  the equations  
( \ref {complex_8F},\ref{complex_aperture} )
 are equivalent ,  as  shown in \cite {Okulov:1988} by  
asymptotic evaluation , to solution 
of  partial differential  MBE  for  class - A  laser ,
 i.e. GLE  (\ref {class_A_laser} ) .  The  GLE had 
been obtained in \cite {Okulov:1988} through  expanding  of  $E$  
in Taylor series up to the second order . 
This is justified when $L$ is sufficiently  
short  and  fast oscillations of  the  
kernel of  (\ref {complex_8F} ) : 
\begin{equation}
\label{kernel}
\exp (\frac{i k (\vec r_{\bot}-\vec r^{'}_{\bot})^2}{2L} 
\end{equation}
quench the integral  for  all   points except  for those, who are located in the 
vicinity of $\vec r_{\bot}$. 

Let  us consider  here  the  more  general  situation, when  
cavity length $L$ is 
 not  so  short and   fourth  order  terms are  to be taken into account. 
In order to 
get from  ( \ref {complex_8F} ) the  evolution equation with  
small  changes from one 
iterate to 
another , let us separate local and nonlocal part of the map 
( \ref {complex_8F} ) by  the following 
substitution \cite {Okulov:1988}:

\begin{eqnarray}
\label{decomposition map}
 E_{n+1}= E_n + \hat {Fr} [E_n] -E_n {\:}+  [f [E_n] -E_n] + O^2 (\alpha)
{\:} , {\:} 
&& \nonumber \\
\alpha \cong \hat {Fr} [E_n] - E_n , [f[E_n]-E_n], 
\end{eqnarray} 
 
The physical  meaning  of this condition is quite natural : the first, 
i.e. zeroth 
-order term $E_n$ is only slightly affected by the second and third terms 
of  ( \ref {decomposition map}) which 
are responsible for diffraction  and  nonlinearity  correspondingly 
 and they are of  
the  same order $\alpha$ , i.e the changes due to nonlocal  and  
local  part  of  map ( \ref {complex_8F} ) 
are  of  equal  weight and additive .  
 The fourth  term is of the order $\alpha^2$  and it 
is insignificant. The second  term of ( \ref {decomposition map}) ,  
i.e. nonlocal  part  of  the  map ( \ref {conv_map}, 
\ref {complex_8F} ) could  be 
evaluated by stationary phase  method  
up  to the forth  order by  means 
of decomposition of E  in Taylor series   in  
the vicinity  of $\vec r_{\bot}$ \cite {Okulov:1988}: 
\begin{equation}
\label{field decomposition}
\ E_n (\vec r_{\bot}-\vec r^{'}_{\bot}) \cong E_n (\vec r_{\bot}) + 
\nabla^2 [E_n (\vec r_{\bot})] \frac{|\vec r_{\bot} - \vec r_{\bot}^{'}|^2} 
{2!} + \nabla^4 [E_n (\vec r_{\bot})] 
\frac{|\vec r_{\bot}-\vec r^{'}_{\bot}|^4}{4!}
\end{equation}
The substitution  of  ( \ref {field decomposition})  
into\ref {complex_8F} )  and  integration over
 $\vec r_{\bot}$ \cite {Okulov:1988} lead  us  to  
evolution  equation provided  time step $n+1$  ,  
$n$ is considered as infinitesimally  
small :  
\begin{equation}
\label{Swift_Hohenberg_filtering}
\frac {\partial E}{\partial t} = \gamma E + \delta E^2 +\beta E^3 
+\eta E^5 + (a+ib)\nabla^2 (E_n\vec r_{\bot}) +
(c+id) \nabla^4 (E_n\vec r_{\bot}) , 
\end{equation}
i.e. complex  Swift - Hohenberg  equation ,  
which  takes  into account  the higher - 
order  spatial  dispersion terms.  
The  above procedure of  derivation  presents  the  
exact  values of  all coefficients in (\ref {Swift_Hohenberg_filtering}) and  
their  connection  with geometrical  
parameters of the cavity . The  nonzero values $c$  and  $d$  , 
arised  here  as a result  
of  spatial  filtering  of  high  transversal   
harmonics  on  diafragm, without the 
inclusion of  finite gain  linewidth  $T_2^{-1}$ . 

\subsection 	{ Renormalization group equation as a universal limit 
for nonlocal map}

The  conceptual progress in  physical  understanding  of the  nonlocal map 
properties  had been made in \cite {Pikovsky:1986}, where 
it was shown that infinite sequence of 
alternating  nonlinear local maps and  
linear nonlocal maps of convolution type behave 
in universal manner. It was shown by renormalization 
group technique, that nonlocal map 
with real kernel :
\begin{equation}
\label{conv_map3}
\ E_{n+1}(\vec{r_2} )= \int  K(\vec{r_1}-\vec{r_2}) f(E_{n}(\vec{r_1})) 
d \vec{r_1}.
\end{equation}
tends to universal form, regardless to peculiarities
 of a  given  spatial  filters  
which form the   nonlinear dispersive  medium:

\begin{equation}
\label{renorm_group_conv_map}
E_{n+1}= \hat G E_n ; \hat G = \exp [\frac {\Delta^2}{2}
\frac {\partial^2}{\partial x^2}] g, 
\end{equation}
where $\Delta$ is the second moment ( deviation )
 of the kernel $K$ , $ \Delta^2= \int K(\vec r)|\vec r|^2 $,  
$g$ is fixed point of  
the  iterates of  the  nonlinear  
local  map  $f$  \cite {Pikovsky:1986}. The (\ref {renorm_group_conv_map})  
is the fixed point of renormalization group equation. 

In the same way we are able to constract renormalization group equation for  
nonlocal  complex  map (\ref {complex_8F}) , 
using its simplifyed version (\ref {decomposition map}) . Really , for small 
longitudinal steps $\Delta L = L /m$  we may use the 
following  asymptotic form of the nonlocal diffraction 
operator : 
\begin{equation}
\label{renorm_group_conv_map}
E_{n+1}= [1+ \frac {i \Delta L}{2k} \Delta_{\bot}] f [E_n] , 
\end{equation}

Considering  only  the weak changes of the field   at  
each  longitudinal  step we get 
in first order : 

\begin{equation}
\label{renorm_group_conv_map2}
E_{n+1}= E_n + [f [E_n]-E_n] +[1+ \frac {i \Delta L}{2k} \Delta_{\bot}][E_n], 
\end{equation}

Because  of the additive form of these map ( \ref {renorm_group_conv_map2} ) , 
we may consider the second and 
third terms of it as  $commutative$  $operators$. 
Thus, after $m$  infinitesimal  steps  $\Delta L =L/m$  
we get resulting nonlocal operator in universal form : 
\begin{equation}
\label{renorm_group_operator}
\hat K_m= [1+ \frac {i \Delta L}{2k} \Delta_{\bot}]^m=
 [1+ \frac {i  L}{2km} \Delta_{\bot}]^m, 
\end{equation}
Taking  into account the formal identity : 
\begin{equation}
\label{renorm_group_operator}
 [1+ \frac {i  L}{2km} \Delta_{\bot}]^m=
[[1+ \frac {i  L}{2km} \Delta_{\bot} ]
^{2km/iL\Delta_{\bot}}]^{iL\Delta_{\bot}/2k}, 
\end{equation}
 	
We  get  the limit of  the  sequence  of  nonlocal operators : 
\begin{equation}
\label{limit_of_operators}
 \lim_{m\rightarrow \infty}
[1+ \frac {i L}{2km} \Delta_{\bot}]^m=
\exp [\frac {i L}{2 k} \Delta_{\bot}], 
\end{equation}
 						
having  in  mind   the  Euler's  limit   for   $e = 2.71828...$ : 
\begin{equation}
\label{limit_of_e_number}
 \lim_{m\rightarrow \infty}
[1+ \frac {1}{m} ]^m=e, 
\end{equation}

Thus the complex renormalization  group equation 
 has  the fixed  point  in  the  form 
of   the  operator exponent : 
 \begin{equation}
\label{limit_of_operators_renorm_final}
E_{n+1}=
\exp [\frac {i L}{2 k} \Delta_{\bot}] f [E_n], 
\end{equation}

\section{Spatially quasi-periodic exact solutions}

	The model of thin nonlinear slice with any translationally symmetric 
$N_0(\vec r)=N_0(\vec r + \vec p)$ gain 
distribution (i.e. homogeneous or spatially periodic) in Fabry-Perot cavity 
\cite {Okulov:1990}  
provides the ample example of exact solutions of the 
eigenfunction problem of the map 
(\ref {complex_8F}) if  the  aperture function has the gaussian form: 

\begin{equation}
\label{complex_aperture}
D(\vec{r_{\bot}} )= D_0 
\exp ( -\frac{ \vec r_{\bot}^2}{2D_a} + i \frac{ k \vec r_{\bot}^2}{2F})
\end{equation}

The first and most general form of solution is in Fourrier series . 
The solution  is spatially periodic with period p which is selected 
by field  to  match the Talbot condition on cavity length   
$L = m p^2/{\lambda}$ \cite {Napartovich:1987,Okulov:1990} :

\begin{eqnarray}
\label{Field_exact11}
E(\vec{r_{\bot}} )=\exp [- \frac {\vec r_{\bot}^2}
{2{D_a}^2(1+i N_f^{-1}+z/F)}
+ i kz]{\:} [1+i N_f^{-1}+z/F]^{-1}
&& \nonumber \\
\sum_{s,l} a_{s,l} exp [\frac{i\pi (sx+ly)-i \pi m(s^2+l^2)}
{1+i N_f^{-1}+z/F}]
{\:}{\:},{\:}{\:}
\end{eqnarray}

where $N_f$ is Fresnel number of 
Talbot  cavity \cite {Okulov:1990} :   
\begin{equation}
\label{Fresnel_number_Talbot}
N_f= \frac{ k {D_{\alpha}}^2}{z} =\frac{ k {D_{\alpha}}^2}{L}=
\frac{ 2 \pi {D_{\alpha}}^2}{m p^2}
\end{equation}

where $m$ is integer , 
showing how many times the half - Talbot length is contained 
within cavity length $L$ . From ( \ref {Fresnel_number_Talbot} ) 
it follows formally, that  $N_f$ 
  is proportional to  $N_p^2$ \cite {Okulov:1993}  , 
 i.e. number  of the periods of  the structure , 
contained within aperture , 
because  $D_{\alpha}=N_p p$ . 
The same connection between these quantities could be obtained from 
purely geometrical considerations, taking 
into account the number of Fresnel zones, placed 
within  $D_{alpha}=N_p p$   ( fig. 2 ) .

The another form of exact solution could be obtained, 
when spatially periodic field 
represented as periodic sequence of Gaussian beams 
which are diffracted on common 
Gaussian diafragm: 
\begin{eqnarray}
\label{Field_exact112}
E(\vec{r_{\bot}} )=\exp [- \frac {\vec r_{\bot}^2}
{2{D_a}^2(1+i N_f^{-1}+z/F)}
+ i kz]{\:} [1+i N_f^{-1}+z/F]^{-1}
&& \nonumber \\
\sum_{m,n}\exp [-\frac{{({\vec r_{\bot}-\vec p_{m,n}})^2}}
{p^2(1+i N_f^{-1}+z/F)}]
{\:}{\:},{\:}{\:}
\end{eqnarray}

A somewhat different exact solution  is obtained in the 
form of periodic arrays of 
Gauss - Laquerre beams. For the first - order 
Gauss -Laquerre beams each is considered 
as elementary optical vortex we are able to
 constract the rectangular  grid  of  
mutually coherent  vortices with opposite topological charges: 

\begin{eqnarray}
\label{Field_exact112}
E(\vec{r_{\bot}} )=\exp [- \frac {\vec r_{\bot}^2}
{2{D_a}^2(1+i N_f^{-1}+z/F)}
+ i kz]{\:} [1+i N_f^{-1}+z/F]^{-1}
&& \nonumber \\ 
\ [ 1 - \sum_{m,n} \exp [ - \frac{({\vec r_{\bot}-\vec p_{m,n}})^2}
{p^2}+ i \phi (-1)^{m+n}{\:}](\vec r_{\bot}^2-p_{m,n}^2)^{1/2} ]
{\:},{\:}{\:}
\end{eqnarray}
 
In the fig.2 the distribution of intensity  and  the  phase of the in-phase vortex 
array shown .  Note the different orientation of "dark lattice" , 
which represent zeros 
of amplitude (i.e. positions of vortices ) , and "bright  lattice",  
which is rotated 
over angle 45 degrees with respect to "dark" one. 
The fig. 3 represents the out-of- 
phase vortex array with antiparallel topological charges. 

\section{Comparison with experiment}

Recent experiments with $L = 2 mm$ long cavity diode-pumped solid-state laser 
\cite {Chen:2001} showed the formation of the quasi-periodic spatial lattices of 
vortices in  the  
near field. The longidutinal mode spacing 
is  $c / ( 2 L n )$  = $50 Ghz$ ( experimentally 
measured value - 60 Ghz \cite {Chen:2001}) . Such cavity length exceeds  
two times   those 
predicted previously \cite {Okulov:1994}, 
although single-frequency lasing 
 and  phase - locking  of 
the 2x2 arrays by Hermite-Gaussian TEM11  -   mode  had 
been obtained readily even for 
the  cavity length $L = 30 mm$ \cite {Kubota:1992}. The transverse size of  
the gain region $D$ was in between  $D = 0.5 - 1.5 mm$ , 
radius of curvature of mirror $R = 50mm$.  
Thus the present microcavity had transverse size $D$ being only somewhat smaller 
( by the factor 3/4 ) than  cavity length $L$ . 
In general situation cavity would exhibit 
nonparaxial dynamics, 
because the partial waves (or rays in geometrical optics approximation) emitted by 
edges of gain region(fig.1) have angle of  tens of degrees 
with the optical axis of the 
cavity. Nevertheless, as it was shown in \cite {Okulov:1993,Okulov:1994} 
such waves 
with very large tilt to 
optical axis do not  survive in the present cavity. The case 
is that they greatly 
increase the threshold of  lasing and  instead of nonparaxial eigenmode  
the set of 
nearly single mode channels of "amplyfiers" is formed , each  having  
the "local" 
Fresnel  number smaller than 
unity  $N_f \sim d^2/(\lambda z_T)$  $< 1$  \cite {Okulov:1993,Okulov:1994}. 
The "global"  Fresnel  number  $N$  of 
 the cavity for $\lambda = 1.064 \mu m$ moved from 100 to 1000.  The most 
interesting observation  to my opinion is that the
number of vortices in array in any direction ( fig.2) in 
\cite {Chen:2001} proved to be in qualitative agreement with 
our earlier prediction  $N_f=N_p^2$   
\cite {Okulov:1990,Okulov:1993,Okulov:1994}. 
The  observed  period of  arrays is equal roughly to 
 $p =\sqrt{L \lambda} \approx 50-60 \mu m$.
The close  period value had been predicted in 
\cite {Okulov:1994} , where transverse  
mode-locking  by  periodic gain  had  been considered.

\section{Multiscale structure of intracavity vortex field}

The interpretation of  the experiment \cite {Chen:2001}  as 
Talbot transverse mode-locking 
could get additional confirmations if one could prove experimentally 
the $p / 2$  shift  
between  intensity patterns on opposite sides of  the cavity 
\cite {Napartovich:1987}. 
When  cavity lengh 
is a half of  Talbot one $L = p^2/\lambda $  the vortices are 
in-phase  (fig. 2,  see also  fig. 3 
from  \cite {Okulov:1994}). The more remarkable experimental evidence 
could be obtained for out  - of 
 - phase Talbot synchronisation. In this case  the 
adjicent vortices have opposite $\pi
- shifted$ phases and  one mirror carries  $p  / 2$  
intensity pattern \cite {Napartovich:1987,Okulov:1994}. 
The 
cavity length in this case is the quarter 
of the Talbot one $L =p^2/2 \lambda   $ 
\cite {Napartovich:1987,Okulov:1994} . This 
feature of Talbot antiphase - locking is 
due to destructive interference of ajaicent 
vortices within cavity , where vortice 
channels are not parallel, but form the bundle 
of intercepting  threads, each with variable width . 
At the distances from the mirrors  
$Lm =   /2 \lambda m$  the sequence  of  tiny arrays having  the  
period  $p_m  = p  / m$  is 
formed. In the fig. 4 the  spatial  layout  of  
intensity between mirrors  is shown. 
The Fourier images of the field in the
 planes  $L_m =  p^2 / 2 \lambda m$  have  the  
period  $p_m  = 
p  / m $.  

\section{Conclusion}
	The nonlocal map approach provides  simple description of  
microchip laser dynamics. We started from conventional 
Maxwell-Bloch equations and 
under approximation of ultrathin  laser cavity we constructed integral 
equation (\ref {conv_map2}), 
which takes into account 
the boundary conditions on cavity mirrors. 
The iterates of such nonlocal map 
are equivalent to time evolution of partial differential equations - 
Maxwell-Bloch equations.
 We got stationary nonlinear solutions of nonlocal map , i.e. 
nonlinear eigenfunctions of high Fresnel number cavity of 
microchip laser.  The 
two-dimensional spatially periodic solutions obtained
 in the form of Fourier series, 
array of zeroth-order Gaussian beams 
and array of first-order Gauss-Laquerre vortices. 
The essential counterpart of 
such periodic structures is Talbot self- imaging, when 
cavity length is in stiff connection with period of the 
transverse structure. We showed 
 the multiscale structure of intracavity 
vortex field when interference pattern in 
different planes of  the cavity have fractional 
periods compared  to  mirror's 
patterns. Our exact solutions contain definite 
connection between Fresnel number 
and number of vortices within laser aperture predicted earlier and observed 
experimentally. Inspite of  very large Fresnel number which offer possibility of 
generating the highly divergent 
nonparaxial waves,  we found that spontaneously arising 
vortex structure acts as highly effective spatial 
filter selecting paraxial waves.

\begin{figure}
\includegraphics[width=8 cm]{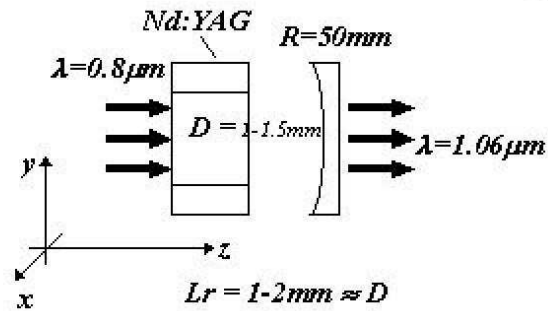}
\caption{Geometry of short length high Fresnel number laser cavity.}
\label{fig.1}
\end{figure}
\newpage 

\begin{figure}
\includegraphics[width=8 cm]{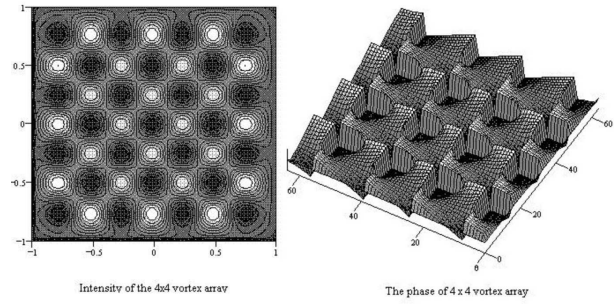}
\caption{Distribution of intensity (left) and 
phase (right) for in-phase vortices array}
\label{fig.2}
\end{figure}
\begin{figure}
\includegraphics[width=8 cm]{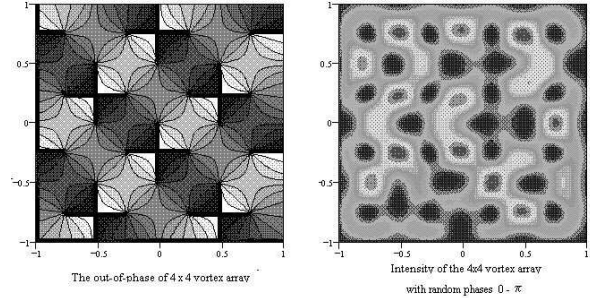}
\caption{Distribution of intensity (left) and 
phase (right) for out-of-phase vortices array}
\label{fig.3} 
\end{figure} 
\begin{figure}
\includegraphics[width=8 cm]{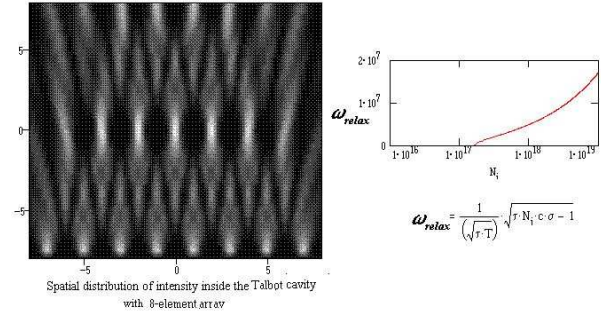}
\caption{Distribution of intensity along microchip 
laser cavity. The 8 in-phase synchronized vortices are shown.
The angular frequency of relaxation of relaxation oscillations 
$\omega_{relax} \sim $ $1/\sqrt {\tau T_1}$ versus density 
of excited $Nd$ ions shown. }
\label{fig.4}
\end{figure}
\end{document}